\documentclass[letterpaper,twocolumn,english,prb,aps,superbib,tightenlines,floatfix]{revtex4-1}
\usepackage[hypertex]{hyperref}
\usepackage{natbib}
\usepackage{graphicx,epsfig}

\begin{document}

\newcommand{\half}{\frac{1}{2}}
\title{Rapid creation of distant entanglement by multiphoton resonant fluorescence}
\author{Guy Z. Cohen} \email{gcohen@physics.ucsd.edu}
\affiliation{Department of Physics, University of California, San
Diego, La Jolla, California 92093-0319}

\author{L. J. Sham}
\affiliation{Department of Physics, University of California, San
Diego, La Jolla, California 92093-0319}

\begin{abstract}
We study a simple, effective and robust method for entangling two
separate stationary quantum dot spin qubits with high fidelity
using multiphoton Gaussian state. The fluorescence signals from
the two dots interfere at a beam splitter. The bosonic nature of
 photons leads, in analogy with the Hong-Ou-Mandel effect, to
 selective pairing of photon holes (photon absences in the fluorescent
signals). As a result, two odd photon number detections at the
outgoing beams herald trion entanglement creation, and subsequent
reduction of the trions to the spin ground states leads to
spin-spin entanglement. The robustness of the Gaussian states is
evidenced by the ability to compensate for photon absorption and
noise by a moderate increase in the number of photons at the
input. We calculate the entanglement generation rate in the ideal,
nonideal and near-ideal detector regimes and find substantial
improvement over single-photon schemes in all
 three regimes. Fast and efficient spin-spin entanglement creation
can form the basis for a scalable quantum dot quantum computing
 network. Our predictions can be tested using current experimental
capabilities.
\end{abstract}

\maketitle

\section{Introduction} \label{introduction}

Quantum computers have the power to solve some problems much more
quickly and efficiently than any classical
computer.\cite{Barnett09,Jaeger07} Quantum communication between
two distant elements in the computer or between two quantum
computers can be established over a quantum channel and used,
e.g., to distribute a key for encrypting data over a classical
channel or to establish entanglement, which can then be used for
quantum teleportation\cite{Bennett93} or dense
coding.\cite{Bennett92} A leading model for the implementation of
these tasks is a quantum network\cite{Kimble08,Luo09} composed of
many nodes, each containing one or several quantum bits (qubits),
and high fidelity quantum channels connecting these nodes.
Preparation, processing and storage of quantum information are
performed at the network nodes, which must therefore be stable and
in fixed locations (stationary qubits), while quantum
communication is attained by photons (flying qubits).

Quantum dot (QD) spin qubits are promising candidates for
stationary qubits in scalable quantum networks, as they are
compatible with existing semiconductor technology, can be
integrated on a chip with photonic crystal
cavities,\cite{Badolato05,Gallo08} have short optical
recombination and photon emission times,\cite{Pelton02,Moreau02}
can be manipulated by fast
single-qubit\cite{Press08,Foletti09,Greilich09,Kim10} and
two-qubit\cite{Brunner11,Kim11,Chen12,Shulman12} quantum gates,
and can be entangled with adjacent qubits by tunneling
interaction\cite{Kim11} or with remote ones via entanglement
swapping with photons.\cite{DeGreve12,Gao12,Schaibley13}

Entanglement of stationary qubits is an essential resource in a
quantum network. Inter-qubit entanglement can be applied to
increase the quantum capacity of a channel,\cite{Bowen02} and to
implement quantum repeaters\cite{Duan01} for long-distance quantum
communication. Entangled qubit pairs can also form the basis for
entanglement-assisted quantum error correction (QEC), which has
fewer constraints and higher capacity than standard
QEC.\cite{Brun06} The heralded entanglement of two remote
stationary qubits can be achieved by first entangling the qubits
with photons, which interfere on a beam splitter, and then
performing a measurement on the photons, with one (type I) or two
(type II) single photon detections indicating whether the
entanglement creation was successful.\cite{Zippilli08,Luo09} Type
I heralded entanglement was suggested
theoretically\cite{Cabrillo99,Childress05} and
implemented,\cite{Matsukevich08,Slodicka13} but suffers from a low
 success rate due to the requirement of weak qubit excitation and due to high sensitivity to fluctuations in the photonic
 phase. Type II heralded entanglement
 schemes,\cite{Simon03,Duan03} which were also experimentally
 realized,\cite{Moehring07,Maunz09,Hofmann12}
are exempt from the weak excitation requirement and, through
two-photon interference, are more robust to noise than type I
schemes. However, both single-photon heralded entanglement types
use single-photon states, which are very sensitive to noise, and
as a result they currently produce entangled pairs at a slow rate
of the order of 0.1 to 10 pairs per
minute.\cite{Moehring07,Matsukevich08,Maunz09,Hofmann12,Slodicka13}
A question is raised whether this performance can be improved by
increasing the number of photons at the input.

Several theoretical proposals to improve the entangled pair
generation rate by multiphoton light included a system where the
matter qubit is placed in a resonant cavity and interacts with
coherent state light via coherent Raman
scattering,\cite{vanLoock06} whereas the light at the beam
splitter output undergoes continuous wave homodyne measurement.
The use of multiphoton light, in this case, does not improve the
results substantially over single-photon schemes, as the loss of a
single photon produces ``which path'' information which leads to
significant decoherence, impairing entanglement creation. Another
theoretical work used twin-Fock or NOON states,\cite{Huang08} but
again produced entanglement creation rates of the same order as
single-photon schemes due to sensitivity to single-photon losses.
Recently, robust multiphoton entanglement creation using coherent
state light was put forward.\cite{Chan13} An entanglement creation
rate of $1.2\times 10^7$ pairs per minute was calculated for QD
spin qubits with this scheme.

Gaussian states form a versatile, robust, powerful, yet simple,
continuous variable alternative to qubits in quantum information
processing.\cite{Weedbrook12,Braunstein05} The single-mode
Gaussian state is, in general, squeezed vacuum rotated and
displaced in phase space, and thus includes coherent states,
squeezed states and thermal states as special
cases.\cite{Weedbrook12} Multimode Gaussian states, in turn, can
exhibit entanglement and be used for quantum
cryptography,\cite{Grosshans02} quantum
teleportation,\cite{Braunstein98} quantum
communication,\cite{Holevo01} quantum computation,\cite{Gu09}
quantum cloning\cite{Cerf00} and quantum dense
coding.\cite{Ralph02} Experimentally, Gaussian states are readily
generated with photons\cite{Jing03,Takahashi10} and manipulated by
many common optical elements, such as beam splitters and
quarter-wave plates, which constitute Gaussian
operations.\cite{Weedbrook12} Gaussian measurements can be
performed by homodyne detection.\cite{DAuria09,Buono10}

Among all Gaussian states, the two-mode Gaussian state ($N=2$), a
simple bipartite continuous variable system, attracted much
research effort in recent years. The separability criterion was
given\cite{Duan00} as well as a closed expression for the
entanglement of formation for symmetric states\cite{Giedke03} and
the logarithmic negativity entanglement measure for all
states.\cite{Vidal02} The purity, von Neumann entropy and mutual
information were also found,\cite{Serafini04} as was the quantum
discord,\cite{Giorda10} a measure of the quantumness of
correlations for a given state.

We present a simple, effective, robust and scalable multiphoton
entanglement generation method driven by two-mode Gaussian state
light. Two fluorescence signals interfere at a beam splitter and
are then subject to projective photon number measurement. The
bosonic statistics of photons gives rise through the
\emph{multiphoton Hong-Ou-Mandel effect}, i.e. the Hong-Ou-Mandel
effect as applied to photon holes in a multiphoton state, to
heralded trion entanglement only upon two odd photon number
measurements. Spin entanglement is obtained by reducing the trion
states to their corresponding spin states through coherent Rabi
rotations in the QDs. In contrast with current single-photon
schemes, wherein an excited $\Lambda$ system must undergo
spontaneous emission for entanglement creation,\cite{Zippilli08}
the evolution of our 4-level QD spin qubits before the measurement
is completely deterministic.

The Gaussian state redundancy provides robustness against noise
that is lacking in single-photon or vacuum states. In our system
this robustness is manifested in the ability to offset photon
absorption, noise and low detection efficiency by a moderate
increase in the mean input photon number. We show the entanglement
generation rate in our Gaussian state scheme is substantially
higher than the ones in single-photon schemes for all three
regimes of ideal, nonideal and near-ideal photon number detectors.
Even though we have made no direct use of the continuous variable
information, the strength of the Gaussian information, we hope
that our results
 from employing the Gaussian states contribute to the viability of
this mode of information processing.

The paper is organized as follows. In Sec.~\ref{section1} we
present the system model and Hamiltonian. We set criteria for the
input photon state, find the state satisfying these criteria and
solve the Hamiltonian exactly for this state. In
Sec.~\ref{section2} we describe the entanglement creation protocol
in detail and derive the conditions and probability for heralded
entanglement when ideal photon number detectors are employed. In
Sec.~\ref{section3} we discuss the effect of decoherence, noise
and nonideal detectors on entanglement creation. We arrive at the
conditions and probability for heralded entanglement in the
nonideal case and analyze the limits and properties of the success
probability. In Sec.~\ref{section4} we consider the problem of
false positive measurements and offer several solutions. Finally,
in Sec.~\ref{conclusions} we discuss the key results and consider
directions for future research.

\section{System Model and Hamiltonian} \label{section1}

\begin{figure}[t!]
\begin{center}
\large\flushleft{\textbf{(a)}}\normalsize\\ \ \\
\includegraphics[width=8.5cm]{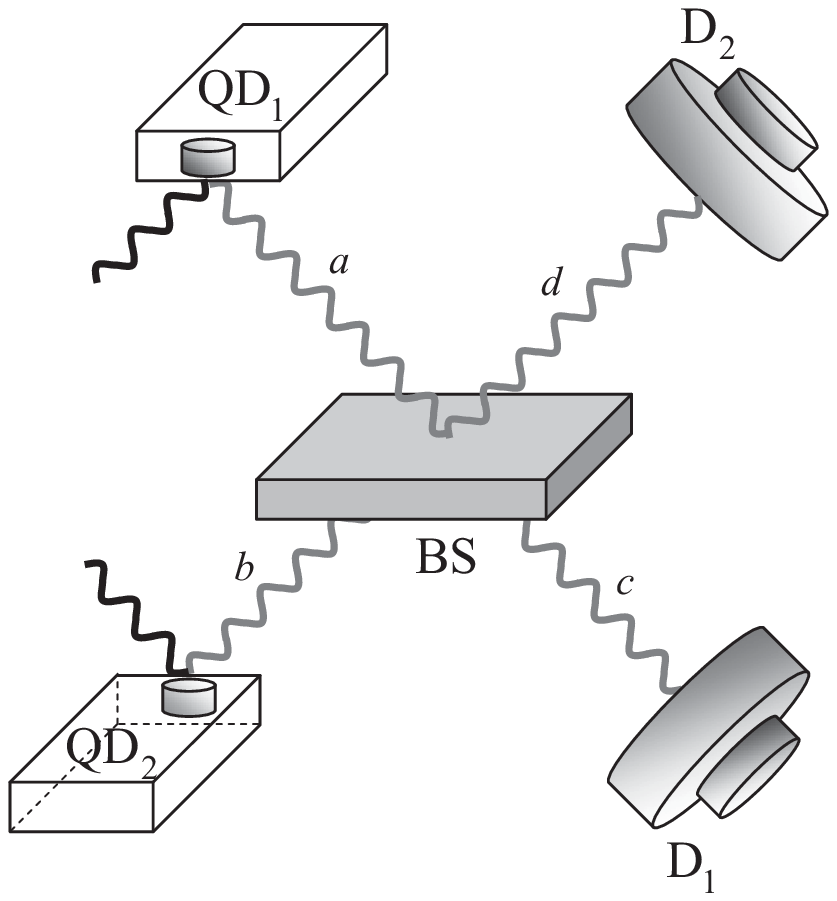}\\
\large\flushleft{\textbf{(b)}}\normalsize
\end{center}
\begin{center}
\includegraphics[width=3cm]{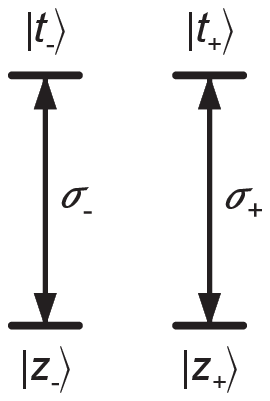}
\end{center}
\caption{\label{fig01} (a) Schematic of the system for
entanglement generation: Two distant quantum dots, $\mathrm{QD}_1$
and $\mathrm{QD}_2$, are driven by coherent light resulting in
fluorescence which interferes at a beam splitter, BS. The outputs
of the beam splitter are detected by two photomultiplier photon
number detectors, $\mathrm{D}_1$ and $\mathrm{D}_2$, which provide
postselection of the entangled state. The photon paths are labeled
$a$, $b$, $c$ and $d$. (b) Energy level diagram for the first dot:
The dot is in the Faraday geometry in the limit of $\mathbf{B}=0$
T, and the allowed transitions are between the $-1/2$ spin state
$|z_-\rangle$ to the $-3/2$ trion state $|t_-\rangle$ and between
the $1/2$ spin state $|z_+\rangle$ to the $3/2$ trion state
$|t_+\rangle$ through $\sigma_\mp$ polarized light, respectively.
The two states $|z_\mp\rangle$ are nearly degenerate, as are
$|t_\mp\rangle$. The energy level diagram for the second dot is
identical except that $z$ and $t$ have tildes.}
\end{figure}
Our system, shown schematically in Fig.~\ref{fig01}(a), consists
of two four-level, singly charged quantum dots (QDs) in the
Faraday geometry\cite{Kim08,Xu08} in the limit of $\mathbf{B}=0$
T, a beam splitter and two photomultiplier photon number
detectors. We note that the zero magnetic field requirement can be
relaxed, so long as the Zeeman-splitting-induced difference in
resonance frequency between the two QD level pairs is much lower
than the trion state linewidth. The QD energy level diagram is
given in Fig.~\ref{fig01}(b). The spin state $\pm 1/2$ of each QD
functions as a qubit, and the goal is to create entanglement
between these two spin qubits. The two levels $|z_\mp\rangle$ are
degenerate, as are $|t_\mp\rangle$. The allowed optical
transitions in the first QD are between the $|z_-\rangle$ and
$|t_-\rangle$ levels through $\sigma_-$ polarized light and
between the $|z_+\rangle$ and the $|t_+\rangle$ levels through
$\sigma_+$ polarized light, while the cross-spin transitions are
forbidden in this geometry. For brevity we take the two QDs to
have the same energy levels and denote the QD$_2$ states by
tildes, $|\tilde{z}_\mp\rangle, |\tilde{t}_\mp\rangle$.

Our protocol for entanglement generation in this system is as
follows. The first and second QDs are initially prepared in the
$|x_+\rangle=(|z_-\rangle+|z_+\rangle)/\sqrt{2}$ and
$|\tilde{x}_+\rangle=(|\tilde{z}_-\rangle+|\tilde{z}_+\rangle)/\sqrt{2}$
states, respectively, by, e.g., a $\pi/2$ laser
pulse.\cite{Atature06} Afterwards, light is shone on each of the
QDs and emerges at the beam splitter inputs with a fluorescence
signal. The photon states at the beam splitter outputs are
detected at the photomultiplier tubes, and the measurement results
herald whether distant entanglement between the qubits was
achieved.

We choose the composite photon state driving the QDs by specifying
a set of conditions which facilitate the entanglement
postselection procedure. (1) The state should interact with both
level pairs $|z_\mp\rangle, |t_\mp\rangle$ in each QD, and hence
it should be a two-mode state, with polarizations $\sigma_+$ and
$\sigma_-$ at the resonance frequency of the levels $\omega_0$.
(2) The state should be a multiphoton Gaussian state, since
Gaussian states are expected to be more robust quantum information
carriers than single-photon states.\cite{Weedbrook12} (3) For
simplicity, we follow the common practice\cite{Moehring07,Maunz09}
in specifying the driving Gaussian state to have the same
parameters for the two polarization modes. (4) The Gaussian state
should be a pure state.

The mathematical details of the construction of the Gaussian state
with the prerequisites defined above is given in Appendix
\ref{app-A}. This state is found to be the EPR state, which in the
number state basis reads
\begin{equation}
|\chi(0)\rangle=\sqrt{1-\lambda^2}\sum\limits_{m=0}^\infty
\lambda^{m} |m,m\rangle\label{eq06},
\end{equation}
where $\lambda=\tanh r$, $r$ is the squeezing parameter, and
$|m_-,m_+\rangle$ are the two-polarization mode number basis
photon states. The average photon number in the EPR state is the
same in both modes,
\begin{equation}
\overline{m}=\frac{\lambda^2}{1-\lambda^2}\label{eq06-2},
\end{equation}
which will be used to characterize the input photon state. The
variances in the photon number, also identical for both modes,
exhibit super-Poissonian statistics, as $\langle (\Delta
m)^2\rangle=\overline{m}^2+\overline{m}$.

The next step is to find the result of the interaction of the
designed input photon state with each QD. We model the process by
the state evolution of the Jaynes-Cummings Hamiltonian
\cite{Jaynes63}
\begin{eqnarray}
H&=&H_0+H_1,\label{eq07}\\
H_0&=&\omega_0\sum\limits_i a_i^\dagger
a_i+\frac{\omega_0}{2}\sum\limits_i \biggl(|t_i\rangle\langle
t_i|-|z_i\rangle\langle z_i|\biggr),\label{eq08}\\
H_1&=&g\sum\limits_i \biggl(a_i^\dagger |z_i\rangle\langle
t_i|+\mathrm{h.c.}\biggr),\label{eq09}
\end{eqnarray}
where $\hbar$ is taken as unity, $g$ is the coupling constant, and
$a_i$ and $a_i^\dagger$ are the annihilation and creation
operators for a $\sigma_i$ polarized photon with the resonance
frequency $\omega_0$. This evolution idealizes the scattering
process of the light against the quantum dot and the host matrix
by assuming it is entirely reflected by the solid system. This
mirror can be approximated by the reflection of the quantum dot
grown on a substrate on top of a metal gate. We attempt here only
a conceptual formulation and recognize the limitation of this
aspect of the estimate of the entanglement efficiency as lacking
in accuracy for an experimental design. The idealization neglects
the photon loss in the actual physical processes. Possible
remedies include wave guides as conduit of photon states and
embedding each QD in a Fabry-P\'{e}rot microcavity tuned to have
modes in resonance with the QD transitions [Fig.~\ref{fig01}(b)].
We provide in the conclusions section an estimate of the errors if
the latter method is employed.

The Hamiltonian in the interaction representation is the time
independent $H_1$ because of the resonance condition.
The state is governed by the Schr\"{o}dinger equation
\begin{equation}
i\frac{\partial}{\partial
t}|\psi(t)\rangle=H_1|\psi(t)\rangle\label{eq11}
\end{equation}
with the initial condition being the product of the QD initial
condition and the photon state in Eq.~(\ref{eq06}),
\begin{equation}
|\psi(0)\rangle=|x_+\rangle|\chi(0)\rangle.\label{eq12}
\end{equation}
The exact solution at time $t$ of Eq.~(\ref{eq11}) with the
Hamiltonian (\ref{eq09}) is
\begin{eqnarray}
|\psi_a(t)\rangle=|x_+\rangle|\chi(t)\rangle+|t_-\rangle|\chi_-(t)\rangle+|t_+\rangle|\chi_+(t)\rangle\label{eq13},
\end{eqnarray}
where the subscript $a$ denotes the photon path in
Fig.~\ref{fig01}, and the photon conditional states are
\vspace{-2em}\begin{center}
\begin{eqnarray}
&\hspace{-4.45em}|\chi(t)\rangle=\sqrt{1-\lambda^2}\sum\limits_{m=0}^\infty\lambda^m
\cos(g\sqrt{m} t)|m,m\rangle,\label{neweq14}\\
&\hspace{-1.5em}|\chi_-(t)\rangle=-i\sqrt{\frac{1-\lambda^2}{2}}\sum\limits_{m=0}^\infty\lambda^m
\sin(g\sqrt{m} t)|m-1,m\rangle,\label{neweq15}\\
&\hspace{-1.5em}|\chi_+(t)\rangle=-i\sqrt{\frac{1-\lambda^2}{2}}\sum\limits_{m=0}^\infty\lambda^m
\sin(g\sqrt{m} t)|m,m-1\rangle.\label{neweq16}
\end{eqnarray}
\end{center}
We see the interaction results in photon and qubit states
entanglement, which will be useful for inter-qubit entanglement
creation. The beam splitter input state is thus
$|\psi_a(t)\rangle|\psi_b(t)\rangle$, with $|\psi_a(t)\rangle$
given in Eq.~(\ref{eq13}) and $|\psi_b(t)\rangle$, the state
corresponding to the $b$ photon path and the second QD, obtained
from the same equation by writing it in terms of
$|\tilde{x}_+\rangle$ and $|\tilde{t}_\pm\rangle$.

\section{Entanglement Generation} \label{section2}

We detail the entanglement process. The input states to the beam
splitter from the two QD's driven by coherent light interfere at
the beam splitter. The total number of photons at each output is
measured by a detector [see Fig.~\ref{fig01}(a)], and an entangled
state is then selected after obtaining the appropriate pair of
photon numbers in the two paths. Let the input composite photon
basis state be $|m_-,m_+;\tilde{m}_-,\tilde{m}_+\rangle_{a;b}$,
which corresponds to $m_-$ and $m_+$ photons of respective
polarizations $\sigma_\mp$ along path $a$ from QD$_1$, and
$\tilde{m}_-,\tilde{m}_+$ photons along path $b$ from QD$_2$. The
output state is a linear combination of the output composite basis
states $|m'_-,m'_+;\tilde{m}'_-,\tilde{m}'_+\rangle_{c;d}$, along
paths $c,d$, defined analogously to the input basis states. The
two basis sets are related by the unitary transformation of the
single photon by the beam splitter,
\begin{eqnarray}
c_i&=&\frac{1}{\sqrt{2}}(a_i+b_i),\nonumber\\
d_i&=&\frac{1}{\sqrt{2}}(b_i-a_i),\label{eq14}
\end{eqnarray}
where $a_i, b_i$, are the annihilation operators of incoming
photons of polarization $i=\mp$ along paths $a,b$, respectively,
and, similarly, $c_i, d_i$ denote the outgoing photons.

The detection process can be described by projection
operators,\cite{Barnett09} each of which corresponds to a
measurement result. Since the photodetection is insensitive to
polarization, the projector for a measurement of $n_1$ and $n_2$
photons in the first and second detectors, respectively, is
\begin{equation}
P_{n_1,n_2}=\sum\limits_{k=0}^{n_1}\sum\limits_{l=0}^{n_2}|k,n_1-k;l,n_2-l\rangle\langle
k,n_1-k;l,n_2-l| \label{eq15}.
\end{equation}
As these projectors are orthonormal, i.e.
$P_{n_1,n_2}P_{n_1',n_2'}=\delta_{n_1,n_1'}\delta_{n_2,n_2'}P_{n_1,n_2}$,
the measurement is a von Neumann measurement, and the state after
the measurement is
\begin{equation}
|\psi_{n_1,n_2}\rangle=P_{n_1,n_2}|\psi(t)\rangle, \label{eq16-1}
\end{equation}
where $|\psi(t)\rangle$ is the state of the system after the beam
splitter and before the measurement. The probability for this
measurement is
\begin{equation}
\mathrm{Prob}_{n_1,n_2}(\overline{m},t)=\langle
\psi(t)|P_{n_1,n_2}|\psi(t)\rangle. \label{eq16-2}
\end{equation}

The measurement probability $\mathrm{Prob}_{n_1,n_2}$ has the same
form in the Schr\"{o}dinger picture as in the interaction picture,
since the transition to the former in Eq.~(\ref{eq16-2}), which
includes insertion of $\exp(iH_0t)$ and $\exp(-iH_0t)$ operators,
leads to the same expression. Thus $\mathrm{Prob}_{n_1,n_2}$,
calculated in the interaction picture, corresponds to the
measurement probability of $n_1$ and $n_2$ photons at time $t$.

Consider first the case when $n_1$ is even and $n_2$ is odd or
vice versa. The state after the measurement is
\begin{equation}
|\psi_{n_1,n_2}\rangle=P_{n_1,n_2}U_{BS}[|\psi_a(t)\rangle|\psi_b(t)\rangle],\label{neweq17}
\end{equation}
where $U_{BS}$ is the beam splitter transformation operator, and
$|\psi_{a,b}(t)\rangle$ are given in Eq.~(\ref{eq13}). The
$P_{n_1,n_2}$ and $U_{BS}$ operators act only on the 9 conditional
photon states in $|\psi_a(t)\rangle|\psi_b(t)\rangle$, and as
$U_{BS}$ conserves the total number of photons, only the terms in
$|\psi_a(t)\rangle|\psi_b(t)\rangle$ with an odd number of photons
contribute. Of these terms $|\chi\rangle|\chi_\mp\rangle$ differs
from $|\chi_\mp\rangle|\chi\rangle$ by $m_\mp\leftrightarrow
\tilde{m}_\mp$ in each composite photon state
$|m_-,m_+;\tilde{m}_-,\tilde{m}_+\rangle_{a;b}$ in it. This
transposition is the same as $a_i\leftrightarrow b_i$, which after
the beam splitter transformation (\ref{eq14}), is equivalent to
$d_i\rightarrow -d_i$ and gives a $(-1)^{n_2}$ overall factor. The
state after the measurement is thus
\begin{equation}
|\psi_{n_1,n_2}\rangle=P_{n_1,n_2}U_{BS}\left
\{|A_-\rangle|\chi\rangle|\chi_-\rangle+|A_+\rangle|\chi\rangle|\chi_+\rangle\right
\},\label{neweq18}
\end{equation}
where
$|A_\mp\rangle=|x_+\rangle|\tilde{t}_\mp\rangle+(-1)^{n_2}|t_\mp\rangle|\tilde{x}_+\rangle$.

The terms $P_{n_1,n_2}U_{BS}(|\chi\rangle|\chi_\mp\rangle)$ in
Eq.~(\ref{neweq18}) have the same norm and are orthogonal, since
$|\chi\rangle|\chi_\mp\rangle$ differ only by an exchange of
polarizations and are orthogonal, a property preserved by
$P_{n_1,n_2}$ and $U_{BS}$. As $|A_-\rangle$ and $|A_+\rangle$
have the same norm and are orthogonal, the state following a $-$
or $+$ measurement at one QD carries no preference for $-$ or $+$
at the other QD, and $|\psi_{n_1,n_2}\rangle$ has no spin
entanglement.

Next, consider the cases where both $n_1$ and $n_2$ are odd or
both $n_1$ and $n_2$ are even. The state after the measurement is
again given by Eq.~(\ref{neweq17}), but now the contributing terms
have an even total photon number. Of these terms
$|\chi_-\rangle|\chi_+\rangle$ differs from
$|\chi_+\rangle|\chi_-\rangle$ by $a_i\leftrightarrow b_i$, which
after the beam splitter transformation (\ref{eq14}), is equivalent
to $d_i\rightarrow -d_i$ and gives an overall sign of $\mp$ in the
odd-odd and even-even cases, respectively. The state after the
measurement reads
\begin{eqnarray}
&|\psi_{n_1,n_2}\rangle=P_{n_1,n_2}U_{BS}\left\{|B_0\rangle|\chi\rangle|\chi\rangle+|B_\downarrow\rangle|\chi_-\rangle|\chi_-\rangle+\right .\nonumber\\
&\left .
|B_\uparrow\rangle|\chi_+\rangle|\chi_+\rangle+|B_\mp\rangle|\chi_-\rangle|\chi_+\rangle\right
\}, \label{neweq21}
\end{eqnarray}
where $|B_0\rangle=|x_+\rangle|\tilde{x}_+\rangle$,
$|B_{\downarrow}\rangle=|t_-\rangle|\tilde{t}_-\rangle$,
$|B_{\uparrow}\rangle=|t_+\rangle|\tilde{t}_+\rangle$, and
$|B_\mp\rangle=(|t_-\rangle|\tilde{t}_+\rangle\mp|t_+\rangle|\tilde{t}_-\rangle)$
are the QD conditional states in the odd-odd and even-even cases,
respectively. The photon states multiplying each of $|B_0\rangle$,
$|B_\downarrow\rangle$ and $|B_{\uparrow}\rangle$ are invariant
under $a_i\leftrightarrow b_i$, which corresponds to
$d_i\rightarrow -d_i$ after the beam splitter transformation. They
therefore contribute only to the even-even case, and the state in
the odd-odd case is $|B_-\rangle
P_{n_1,n_2}U_{BS}(|\chi_-\rangle|\chi_+\rangle)$, separable to a
photon state and qubit state of
$|\Psi_-\rangle=\frac{1}{\sqrt{2}}(|t_-\rangle|\tilde{t}_+\rangle-|t_+\rangle|\tilde{t}_-\rangle)$,
Bell's fourth state, which is maximally entangled. By the
selection rules in Fig.~\ref{fig01}(b), each trion state decays
spontaneously to the corresponding spin state with a random phase.
To preserve the entanglement and avoid the delay, the trion states
should instead be reduced to the spin states by applying on each
QD a pair of phase-locked $\pi$-pulses with polarizations
$\sigma_\mp$.

Having shown that an odd-odd measurement heralds QD spin
entanglement, we wish to investigate the even-even case. The QD
density matrix, found from Eq.~(\ref{neweq21}) by tracing over the
photons in $|\psi_{n_1,n_2}\rangle\langle\psi_{n_1,n_2}|$, is in
this case, up to normalization,
\begin{equation}
\rho=q_0| B_0\rangle\langle B_0|+q_{\updownarrow}(|
B_\downarrow\rangle\langle
 B_\downarrow|+| B_\uparrow\rangle\langle
 B_\uparrow|)+q_+| B_+\rangle\langle B_+|,\label{neweq22}
\end{equation}
where the conservation of total photon number in each polarization
was used to show that the photon states associated with
$|B_0\rangle$, $|B_\downarrow\rangle$, $|B_\uparrow\rangle$, and
$|B_+\rangle$ in Eq.~(\ref{neweq21}) do not lead to cross terms in
the QD density matrix. The terms $|B_\downarrow\rangle\langle
 B_\downarrow|$ and $|B_\uparrow\rangle\langle
 B_\uparrow|$ have the same coefficient in
Eq.~(\ref{neweq22}), since the photon states associated with
$|B_\downarrow\rangle$ and $|B_\uparrow\rangle$ in
Eq.~(\ref{neweq21}) are identical apart from an exchange of the
polarizations. The QD density matrix in Eq.~(\ref{neweq22}) is a
mixture of the density matrix $q_0| B_0\rangle\langle B_0|$, which
is separable by the form of $|B_0\rangle$, and the density matrix
given by the rest of Eq.~(\ref{neweq22}). The last density matrix
has $q_{\updownarrow}\geq q_+$, as shown in Appendix \ref{app-B},
and thus has zero concurrence.\cite{Wootters98} We conclude we
have no spin entanglement in the even-even case, as the density
matrix in Eq.~(\ref{neweq22}) is separable, being a mixture of two
separable density matrices.\cite{Barnett09}

The association of spin entanglement with odd-odd measurements can
also be seen to emerge from the bosonic nature of photons. The
interaction with the QDs results in coupling photon hole states to
$|t_-\rangle$ and $|t_+\rangle$ in Eq.~(\ref{eq13}). The part of
$|\psi_a(t)\rangle|\psi_b(t)\rangle$ with no photon holes has an
even total photon number and a photon state symmetric under
$a_i\leftrightarrow b_i$ and hence must produce only even-even
measurements. The parts with a single photon hole have an odd
total photon number and contribute to odd-even or even-odd
measurements. The parts involving two photon holes with the same
polarization are associated with
$|t_\mp\rangle|\tilde{t}_\mp\rangle$. For these parts, the photon
holes interfere, owing to their bosonic statistics, \`{a} la the
Hong-Ou-Mandel (HOM) effect\cite{Hong87} to give a pair of holes
at either one of the detectors, keeping the even-even measurement.
This effect may be termed the \emph{multiphoton HOM effect}, as
photon holes exist in a background multiphoton state. The parts
involving two photon holes with different polarizations are
associated with $|t_\mp\rangle|\tilde{t}_\pm\rangle$. The photon
holes evolve independently then and can either end up at the same
detector or separate to give an odd-odd measurement. The photon
state contributing to this measurement is the one antisymmetric
under $a_i\leftrightarrow b_i$. Since the associated qubit state
is $|\Psi_-\rangle$, an odd-odd measurement heralds spin
entanglement.

Having dealt with the states after the measurements, we now
analyze the probability to obtain a measurement result,
$\mathrm{Prob}_{n_1,n_2}$ in Eq.~(\ref{eq16-2}). First, since the
beam splitter inputs are identical,
\begin{equation}
\mathrm{Prob}_{n_1,n_2}(\overline{m},t)=\mathrm{Prob}_{n_2,n_1}(\overline{m},t).
\label{eq22}
\end{equation}
We are only interested in the odd-odd case for
$\mathrm{Prob}_{n_1,n_2}$, as this is when entanglement is
generated. By Eq.~(\ref{eq16-2})
$\mathrm{Prob}_{n_1,n_2}=|P_{n_1,n_2}|\psi(t)\rangle|^2$. The
discussion following Eq.~(\ref{neweq21}) points to the term
multiplying $|B_-\rangle$ as the only one remaining in
$|\psi_{n_1,n_2}\rangle$ in the odd-odd case, whereupon by
Eq.~(\ref{neweq21})
\begin{eqnarray}
&\mathrm{Prob}_{n_1,n_2}=\left
|P_{n_1,n_2}U_{BS}|\chi_-\rangle|\chi_+\rangle\right |^2.
\label{neweq23}
\end{eqnarray}

Plugging in $P_{n_1,n_2}$ from Eq.~(\ref{eq15}) in
Eq.~(\ref{neweq23}), using total photon number conservation, and
the beam splitter matrix elements being real, we obtain
\begin{eqnarray}
&\mathrm{Prob}_{n_1,n_2}=\half
(1-\lambda^2)^2\lambda^{2(n_3+1)}\sum\limits_{k=0}^{n_1}\sum\limits_{l=0}^{n_2}\times\nonumber\\
&\left
[\sum\limits_{m=0}^{n_3}\sin(g\sqrt{m}t)\sin(g\sqrt{m'}t)F_{klmm'}\right
]^2, \label{neweq24}
\end{eqnarray}
where $n_3=(n_1+n_2)/2$, $m'=n_3+1-m$, and $F_{klmm'}=\langle
k,n_1-k;l,n_2-l|U_{BS}|m-1,m;m',m'-1\rangle$. The measurement
probability in Eq.~(\ref{neweq24}) is seen to be separable to an
$\overline{m}$-dependent part and a time dependent part. With the
aid of Eq.~(\ref{eq06-2}) we have
$\mathrm{Prob}_{n_1,n_2}(\overline{m},t)=f_{n_1,n_2}(\overline{m})g_{n_1,n_2}(t)$
with
\begin{equation}
f_{n_1,n_2}(\overline{m})=\frac{1}{(\overline{m}+1)^2}\left
(\frac{\overline{m}}{\overline{m}+1}\right )^{n_3+1} \label{eq24}
\end{equation}
and
\begin{eqnarray}
&g_{n_1,n_2}(t)=\half\sum\limits_{k=0}^{n_1}\sum\limits_{l=0}^{n_2}\times\nonumber\\
&\left
[\sum\limits_{m=0}^{n_3}\sin(g\sqrt{m}t)\sin(g\sqrt{m'}t)F_{klmm'}\right
]^2.\label{eq26}
\end{eqnarray}

\begin{figure}[t!]
\begin{center}
\includegraphics[width=8.5cm]{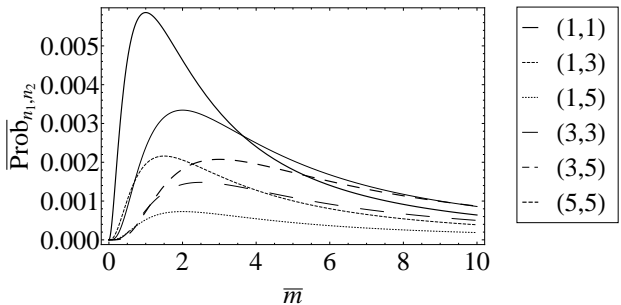}\\
\end{center}
\caption{\label{fig02} Time-averaged probability for measuring
$(n_1, n_2)$ in detectors D$_1$ and D$_2$, respectively, as a
function of the average input photon number $\bar{m}$ for selected
odd-odd pairs of $(n_1, n_2)$. }
\end{figure}

The two time scales in the Hamiltonian (\ref{eq07}) are the
reciprocals of the light frequency $\omega_0$ and the coupling
constant $g$. If the time interval between detections is much
higher than $\frac{2\pi}{g}$, which is
typically\cite{Xu08,Berman11} of the order of 1ns, then for
$g_{n_1,n_2}(t)$ in Eq.~(\ref{eq26}) only the time average is of
practical interest. Let us denote this average as $C_{n_1,n_2}$
and the time-averaged probability of measurement as
\begin{equation}
\overline{\mathrm{Prob}}_{n_1,n_2}(\overline{m})=\frac{C_{n_1,n_2}}{(\overline{m}+1)^2}\left
(\frac{\overline{m}}{\overline{m}+1}\right )^{n_3+1},\label{eq31}
\end{equation}
where Eq.~(\ref{eq24}) was used. The form of $C_{n_1,n_2}$ is
found in Appendix \ref{app-C} and used to plot
$\overline{\mathrm{Prob}}_{n_1,n_2}(\overline{m})$ for several
values of $n_1$ and $n_2$ in Fig.~\ref{fig02}. The maximum of this
function occurs at $\overline{m}=(n_3+1)/2$.

The probability of both detectors measuring an odd number of
photons is
\begin{equation}
\mathrm{ProbSucc}(\overline{m},\eta=1)=
\sum\limits_{n_1=1,3,\dots}^\infty\sum\limits_{n_2=1,3,\dots}^\infty\overline{\mathrm{Prob}}_{n_1,n_2}(\overline{m}),\label{eq34}
\end{equation}
 Substitution of Eq.~(\ref{eq31}) in Eq.~(\ref{eq34}) leads to
 \begin{eqnarray}
\mathrm{ProbSucc}(\overline{m},\eta=1) =
 \frac{\sum\limits_{n_3=1}^\infty D_{n_3}
 \left(\frac{\overline{m}}{\overline{m}+1}\right )^{n_3+1}}{(\overline{m}+1)^2}, \label{eq35}
\end{eqnarray}
where
\begin{equation}
D_{n_3} = \sum\limits_{n_1=1,3,\dots}^{n_3}C_{n_1,2n_3-n_1}
=\frac{n_3}{16}+\frac{1-(-1)^{n_3}}{64},\label{eq36}
\end{equation}
the sum over $n_1$ being evaluated using Eq.~(\ref{eq32}).
Equation (\ref{eq35}) reduces to the simple form,
\begin{equation}
\mathrm{ProbSucc}(\overline{m},\eta=1)=\frac{1}{32}\frac{4\overline{m}+3}{2\overline{m}+1}\left
(\frac{\overline{m}}{\overline{m}+1}\right )^2\label{eq37}
\end{equation}
plotted in Fig.~\ref{fig05}. The asymptotic success probability at
$\overline{m}\rightarrow \infty$ is seen by Eq.~(\ref{eq37}) to be
$1/16$. Since the success probability is monotonously increasing,
we define the characteristic value of $\overline{m}$ needed to
obtain entanglement, $\overline{m}_{1/2}$, as the one giving half
this probability. This value for ideal detectors is, by
Eq.~(\ref{eq37}), $\overline{m}_{1/2}\approx 2.08$.

\section{Impact of Noise} \label{section3}

We will now discuss possible sources of noise which affect the
entanglement generation.
 The important factor is the effective detector efficiency $\eta$, which is the product of optical
transmission factors, detector light collection solid angle ratio
and detector quantum efficiency. The noise from imperfect
generation of the EPR state is negligible, as this state can be
prepared with high fidelity.\cite{Wenger05,Ourjoumtsev06} The
noise from imperfect beam splitter and photon loss in the medium
can be dealt with by introducing a constant multiplicative factor
at the effective detector efficiency
$\eta$.\cite{Ourjoumtsev06,Moehring07} In addition, current
typical detector dark count rates\cite{DeGreve12} of $\sim$ 100 Hz
are much lower than the entanglement creation rate, and the spin
qubit decoherence time of $\sim$ 1 $\mu$s, attainable by nuclear
spin quieting techniques,\cite{Xu09,Press10} is much larger than
the time needed to generate entanglement following qubit state
preparation.

The probability to measure $m$ photons using a detector of
efficiency $\eta$ is given by the quantum theory of
photodetection\cite{Berman11} as
\begin{equation}
\overline{P}_m=\sum\limits_{n=m}^\infty \overline{P}_n{n\choose
m}\eta^m(1-\eta)^{n-m},\label{eq38}
\end{equation}
where $\overline{P}_n$ is the probability for $n$ photons at the
detector input state. For a number state of $q$ photons,
$\overline{P}_n=\delta_{q,n}$, and the probability to measure zero
photons is, by Eq.~(\ref{eq38}), $\overline{P}_0=(1-\eta)^q$. This
is the important probability for us, since measuring zero photons
in either one of the detectors indicates that either the number of
photons is zero, for which we get no entanglement, or that the
detector did not interact with the light at all.

With nonideal detectors\cite{Moehring07,Maunz09} $\eta\ll 1$, and
discerning the odd-odd measurement results from the rest is not
realistic. We therefore consider coincident clicks in both
detectors as a positive result with the understanding that some
false positive readings will occur. With this understanding, the
success probability for nonideal detectors is
\begin{eqnarray}
&\hspace{-1.5em}\mathrm{ProbSucc}(\overline{m},\eta)=\nonumber\\
&\hspace{-1.5em}\sum\limits_{n_1}\sum\limits_{n_2}\overline{\mathrm{Prob}}_{n_1,n_2}(\overline{m})(1-(1-\eta)^{n_1})(1-(1-\eta)^{n_2}),\label{eq39}
\end{eqnarray}
where the sum over $n_1$ and $n_2$ is, as usual, only over the odd
integers, and $\overline{\mathrm{Prob}}_{n_1,n_2}(\overline{m})$
is given in Eq.~(\ref{eq31}). Equation (\ref{eq39}) reduces to
Eq.~(\ref{eq34}) in the ideal detector limit of $\eta=1$. For
other values of $\eta$ Eq.~(\ref{eq39}) is evaluated in Appendix
\ref{app-D} and plotted in Fig.~\ref{fig05}. We find the
asymptotic value of the success probability is still $1/16$ as in
the ideal case, but higher values of $\overline{m}$ are needed to
obtain a given success probability.

\begin{figure}[t!]
\begin{center}
\includegraphics[width=8.5cm]{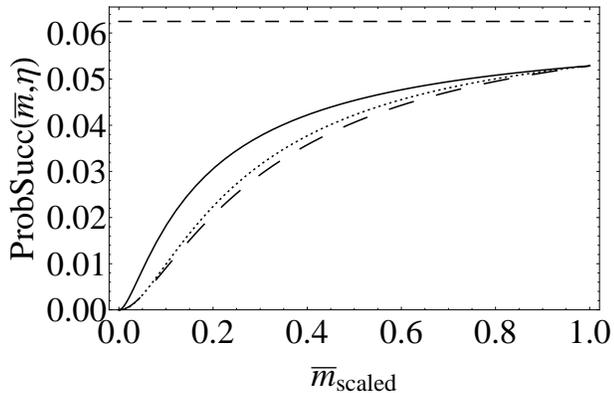}\\
\end{center}
\caption{\label{fig05} Success probability of entanglement
creation vs the scaled average number of photons at the input EPR
state when the effective detector efficiency is $\eta=1,0.1,0.01$
and
$\overline{m}_{\mathrm{scaled}}=\overline{m}/10,\overline{m}/33.3,\overline{m}/251$,
respectively (solid, dotted, large dashes). The average number of
photons at the input EPR state is $\overline{m}$, and the scaling
was chosen so as to make the value at
$\overline{m}_{\mathrm{scaled}}=1$ identical for the three values
of $\eta$. The line for $\eta=0.001$ with the scaling of
$\overline{m}_{\mathrm{scaled}}=\overline{m}/2440$ was not plotted
as it coincided with the line for $\eta=0.01$ to within $0.5\%$.
The reason for the coincidence is the scaling law, which applies
at $\eta\ll 1$ and gives the success probability
$\mathrm{ProbSucc}(\overline{m},\eta)=\mathrm{ProbSucc}(\lambda\overline{m},\lambda^{-1}\eta)$,
where $\lambda$ is the scaling constant. The horizontal dashed
curve shows the asymptotic value of $1/16$ for the probability at
$\overline{m}\rightarrow\infty$, which is the same for all values
of $\eta$.}
\end{figure}

Consider now the behavior of the
$\mathrm{ProbSucc}(\overline{m},\eta)$ function at various limits.
First, at $\overline{m}\ll 1$, we see from Eq.~(\ref{eq31}) that
the dominant term in Eq.~(\ref{eq39}) is the one with $n_1=n_2=1$.
In this limit we thus have
\begin{equation}
\mathrm{ProbSucc}(\overline{m},\eta)\approx
\frac{3}{32}(\overline{m}\eta)^2.\label{eq41}
\end{equation}
The asymptotic behavior at $\overline{m}\rightarrow \infty$ can be
found by the approximations in Appendix \ref{app-D}, which, as
noted there, are exact at this limit. We find
\begin{equation}
\mathrm{ProbSucc}(\overline{m},\eta)\approx
\frac{1}{16}-\frac{1}{\overline{m}}\left [\frac{7 \eta ^2-6 \eta
-8}{4 \eta(\eta -2)}\right ].\label{eq42}
\end{equation}

\begin{figure}[t!]
\begin{center}
\includegraphics[width=8.5cm]{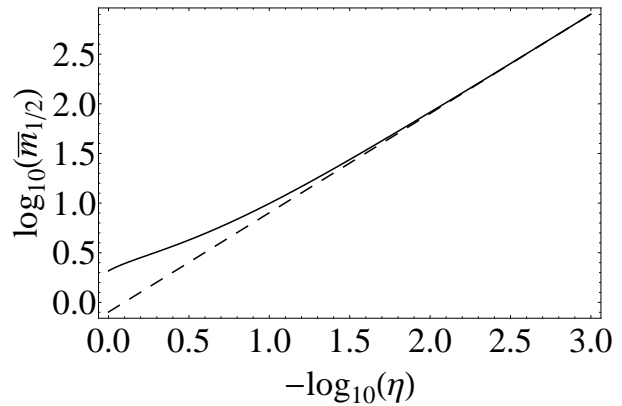}\\
\end{center}
\caption{\label{fig06} Log-log plot of the characteristic average
photon number of the EPR state needed to obtain entanglement,
$\overline{m}_{1/2}$, vs the effective detector efficiency,
$\eta$, showing the range of validity for scaling. The dashed line
corresponds to the asymptotic scaling law of
$\overline{m}_{1/2}=0.80/\eta$.}
\end{figure}

When $\eta\ll 1$ Eq.~(\ref{eq42}) reduces to
\begin{equation}
\mathrm{ProbSucc}(\overline{m},\eta)\approx
\frac{1}{16}-\frac{1}{\overline{m}\eta},\label{eq43-1}
\end{equation}
which is a function of $\overline{m}\eta$, as was the case in
Eq.~(\ref{eq41}) for $\overline{m}\ll 1$. By the approximations in
Appendix \ref{app-D} we find this dependence holds for all values
of $\overline{m}$ when $\eta\ll 1$, and in particular for
$\overline{m}_{1/2}$, defined above by
$\mathrm{ProbSucc}(\overline{m}_{1/2},\eta)=1/32$, as shown in
Fig.~\ref{fig06}. Consequently, the success probability satisfies
the \emph{scaling law}\cite{Chaikin95}
\begin{equation}
\mathrm{ProbSucc}(\overline{m},\eta)=\mathrm{ProbSucc}(\lambda\overline{m},\lambda^{-1}\eta),\label{eq43}
\end{equation}
where $\lambda$ is a constant. This scaling law means that
increasing the average EPR state photon number by a factor of
$\lambda$ is equivalent to improving the detector efficiency by
the same factor. The dependence of the
$\mathrm{ProbSucc}(\overline{m},\eta)$ function on a single
variable $\overline{m}\eta$ enables us to calculate the value of
the function for arbitrary values of $\eta$ or $\overline{m}$, so
long as scaling applies, by its values for a single value of
$\overline{m}$ or $\eta$. This property is called \emph{data
collapse}.\cite{Chaikin95}

\section{False Positive Readings} \label{section4}
The entanglement creation success probability was seen to reach
the same asymptotic value of $1/16$ when the mean input photon
number $\overline{m}$ is large enough to compensate for the low
detector efficiency. However, this success probability is merely
the probability of true positive measurements, while the
probability for false positive measurements may be significant due
to the inability to discern the parity of the photon number with
low efficiency detectors. The probability to obtain false positive
results equals the total positive measurement probability, the
probability for at least one click at each detector, minus the
true positive probability. It is found in analogy with the
derivation leading to Eq.~(\ref{eq39}) and is
\begin{eqnarray}
&\mathrm{ProbFP}(\overline{m},\eta)=\sum\limits_{n_1=1}^\infty\sum\limits_{n_2=1}^\infty\overline{\mathrm{Prob}}_{n_1,n_2}(\overline{m})\times\nonumber\\
&[1-(1-\eta)^{n_1}][1-(1-\eta)^{n_2}]-\mathrm{ProbSucc}(\overline{m},\eta).\label{eq44}
\end{eqnarray}
By noting that $\overline{\mathrm{Prob}}_{n_1,n_2}$ varies
relatively smoothly as a function of its arguments, one finds the
false positive probability to be of the same order of magnitude as
the true positive probability.

There are many ways to cope with false positive measurements. One
method is entanglement distillation through local operations and
classical communications (LOCC) of all the positive measurement QD
pairs obtained.\cite{Bennett96} Other methods include performing a
Bell measurement on the two QDs or unambiguous state
discrimination via the projectors:
$\widehat{\pi}_1=|\Psi_-\rangle\langle\Psi_-|$,
$\widehat{\pi}_2=1-|\Psi_-\rangle\langle\Psi_-|$. This projective
measurement enables us to pick out all the Bell state pairs
without affecting their states.\cite{Barnett09}

Currently, the two aforementioned methods are not available for QD
spin qubits, and hence we suggest avoiding the false positive
readings problem by using low-noise high-efficiency
photon-number-resolving detectors. In recent years, such detectors
with efficiencies in the range of 90-95\% were
developed.\cite{Rosenberg05,Khoury06,Lita08} If we use detectors
of high efficiency in our scheme and can keep the optical
transmission factors high, leading to an effective detector
efficiency above 80\%, then using $\overline{m}=5$ will enable us
to discern between even and odd number of photons, largely
eliminate the false positive counts and obtain close to asymptotic
performance. In comparison, the low entanglement generation rate
in single-photon schemes entails a higher ratio of false positive
to true positive measurements due to detector dark counts, and
this, in turn, translates to a lower entanglement
fidelity.\cite{Blinov04,Moehring07}

\begin{figure}[t!]
\begin{center}
\includegraphics[width=8.5cm]{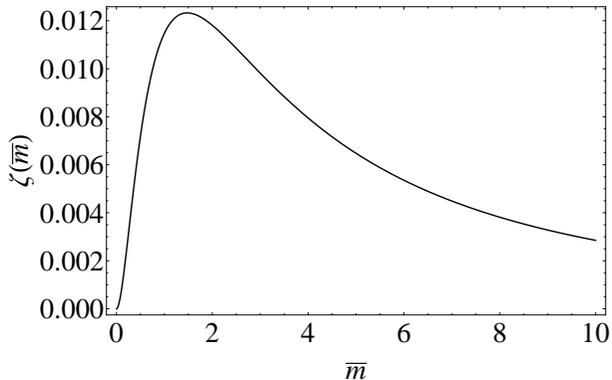}\\
\end{center}
\caption{\label{fig07} Plot of $\zeta(\overline{m})$, the
coefficient of the first order correction in $(1-\eta)$ to the
success probability of an ideal detector, vs $\overline{m}$, the
average photon number at the input EPR state. The effective
detector efficiency, $\eta$, is assumed to be close to 1. The
maximum occurs for $\overline{m}=1.47$ and is $0.0123$.}
\end{figure}

We now wish to show that our scheme outperforms single-photon
schemes also in this near-ideal detector regime. Even though the
success probability for some single-photon
schemes\cite{Moehring07,Olmschenk09} is $\eta^2/4$, which leads to
an asymptotic $\eta=1$ limit of $1/4$, compared with $1/16$ in our
scheme, the single-photon schemes suffer from a factor of the
order of $0.1$ in $\eta$, which currently cannot be eliminated,
due to nonideal single-photon sources. Moreover, the multiphoton
state is more robust against noise, due to the fluorescence signal
being redundantly encoded in multiple components of the state as
can be qualitatively seen from Eq.~(\ref{eq13}). This claim is
quantitatively confirmed by writing the success probability as a
series in the small parameter $(1-\eta)$,
\begin{eqnarray}
&\mathrm{ProbSucc}(\overline{m},\eta)=\mathrm{ProbSucc}(\overline{m},1)-\zeta(\overline{m})(1-\eta)\nonumber\\
&+O((1-\eta)^2).\label{eq45}
\end{eqnarray}
While for the single-photon schemes, $\zeta(\overline{m})=1/8$,
for our scheme $\zeta(\overline{m})$, found numerically from
Eq.~(\ref{eq39}), is plotted in Fig.~\ref{fig07} and is bound from
above by 0.0123 which is more than 10 times lower than the
single-photon result. For $\overline{m}=5$ suggested above this
ratio is even greater.

\section{Conclusions} \label{conclusions}
In this work, we demonstrated a simple, effective and robust
protocol for generating entanglement between two stationary QD
spin qubits using multiphoton Gaussian states. Our method is
clearly distinguished from current single-photon schemes by the
absence of fragile single-photon states and noisy vacuum states
and by unitary evolution before measurement in a spin qubit
4-level system, rather than spontaneous emission in a 3-level one.
We showed resonance fluorescence entails entanglement between the
photon and qubit states, and that measurement of photon numbers
can herald a maximally entangled state of the QDs. The association
of this state with two odd measurements was observed to be a
consequence of the bosonic nature of photons via the
\emph{multiphoton Hong-Ou-Mandel effect}. For nonideal optical
elements and photon number detectors, we pointed that, owing to
the robustness to noise of the Gaussian states, the same success
probability as in the ideal case can be obtained by an increase of
the number of photons at the input. We then proposed the recently
developed near-ideal detectors as means for decreasing false
positive readings at the photon detectors, and found that also in
this regime Gaussian states make our system less prone to noise
than parallel single-photon schemes.

Throughout the paper we assumed the two QDs are identical, as are
the two input EPR states, whereas, in practice, the parameter
values may differ. We now show the results are maintained if the
discrepancy is not too large. First, a resonance frequency
difference between the QDs does not change the results if it is
much lower than the trion state linewidth. Second, if the input
EPR states parameters are $\lambda$ and $\lambda+\Delta\lambda$,
and the QD coupling constants are $g$ and $g+\Delta g$, then with
$\overline{m}>2$, $\Delta \lambda/\lambda<0.01\%$ and $\Delta
g/g<0.01\%$, the expected fidelity of the qubit state relative to
$|\Psi^-\rangle$ upon two odd photon number measurements at
$0<t<40\pi/g$ is found to be higher than 99.7\%.

With the caveats given following Eq.~(\ref{eq07}) concerning the
system presented here not being a fully fledged experimental
design, we hereby estimate the entanglement creation rate. The
ideal photon process is now assumed to take place with QDs in
microcavities and thus exhibits losses, which are now estimated.
The strong coupling regime, which was experimentally realized in
microcavities,\cite{Ates09,Loffler05} is defined by
$\Gamma_{cav}/2<\Omega$,\cite{Grynberg10} where
$\Gamma_{cav}=\omega/Q$ is the cavity decay rate, $\omega$ is the
radiation frequency, $Q$ is the cavity quality factor, and
$\Omega$ is the Rabi frequency. In this regime, assumed here, Rabi
oscillations have a decay rate $\Gamma_{cav}$. The present
collection efficiency of light out of a planar microcavity
containing a QD can be made to be better than
10\%.\cite{DeGreve12,Ates09,Gao12} The probability for resonance
fluorescence in the microcavity, in turn, can be made to be higher
than 50\%.\cite{Ates09} By taking these lower limits, the
probability for resonance fluorescence and successful collection
from each quantum dot is $C\approx 5\%$. QD spin qubit state
initialization in the Faraday geometry was performed within $\sim$
1 $\mu$s,\cite{Atature06} while EPR states were prepared in $\sim$
1.3 $\mu$s.\cite{Wenger05,Ourjoumtsev06} With photodetection time
being much shorter,\cite{DeGreve12} the repetition rate $R$ is
approximately the reciprocal of the larger of these two times. If
we now assume that the average number of photons in the Gaussian
states is large enough for the success probability to reach its
characteristic value of $P=1/32$, the entanglement generation rate
is $C^2PR\approx 3.6\cdot 10^3$ min${}^{-1}$, which is 3 orders of
magnitude larger than the single-photon heralded entanglement
creation
rate.\cite{Moehring07,Matsukevich08,Maunz09,Hofmann12,Slodicka13}
In addition, the spin qubit entanglement fidelity may be expected
to be close to unity assuming high-fidelity preparations of the QD
states and Gaussian photon states.

The experimental realization of this proposed entanglement method
may be an important step towards fast and scalable quantum
computation and communication. More generally, our work provides
one illustration of Gaussian photon states being a viable
alternative to single-photon ones as information carriers. In
analogy, future work may be the application of spin Gaussian
states to replace single-spin qubits as information carriers.
Another direction may be to explore the use of spin qubits to
influence photon Gaussian states for Gaussian information
processing. Our method also has a limitation. The selection rules
in the Faraday geometry provide excellent insulation of two-photon
state evolution paths, which can be utilized for distant
entanglement of stationary qubits. The Voigt geometry has many
advantages, such as fast initial state preparation,\cite{Xu07} but
our entanglement method will have to be drastically redesigned in
this geometry, as its selection rules allow transitions between
each spin state and both trion states.

\appendix
\section{Gaussian States}
\label{app-A} We provide the detail for writing down the input
Gaussian state with conditions specified in Sec.~\ref{section1}. A
Gaussian state can be defined via its Wigner quasiprobability
distribution, given by\cite{Weedbrook12}
\begin{equation}
W(\mathbf{x})=\frac{\exp[-(1/2)(\mathbf{x}-\overline{\mathbf{x}})^\mathrm{T}\mathbf{V}^{-1}(\mathbf{x}-\overline{\mathbf{x}})]}{(2\pi)^N\sqrt{\mathrm{det}~\mathbf{V}}},\label{eq00}
\end{equation}
where
\begin{eqnarray}
\mathbf{x} &=&(q_1,p_1,q_2,p_2,\dots,q_N,p_N), \\
\overline{\mathbf{x}} &=& (\overline{q}_1,\overline{p}_1,\overline{q}_2,\overline{p}_2,\dots,\overline{q}_N,\overline{p}_N), \\
V_{ij} &=& \half \langle
\widehat{x}_i\widehat{x}_j+\widehat{x}_j\widehat{x}_i\rangle-\langle
\widehat{x}_i\rangle\langle \widehat{x}_j\rangle,
\end{eqnarray}
 are the phase
space coordinates vector, displacement vector and covariance
matrix, respectively, with $N$ being the number of modes. The
phase space operators are given by
$\widehat{q}_k=(a_k+a_k^\dagger)/\sqrt{2}$ and
$\widehat{p}_k=-i(a_k-a_k^\dagger)/\sqrt{2}$, where $\hbar$ is put to unity for
brevity.

The Gaussian state is characterized completely by its displacement
vector $\overline{\mathbf{x}}$ and covariance matrix $V$. A
general two-mode Gaussian state can be reduced by local unitary
operations on each mode, which do not affect the purity and
entanglement in the state, to the standard form,\cite{Duan00}
where $\mathbf{x}=0$ and
\begin{equation}
V=\left ( \begin{array}{cccc} s_- & 0 & h_- & 0 \\ 0 & s_- & 0 &
h_+
\\ h_- & 0 & s_+ & 0 \\ 0 & h_+ & 0 & s_+ \end{array} \right )\label{eq01}.
\end{equation}
Condition (3) of the same parametric strength of the two
polarization modes in Sec.~\ref{section1} is satisfied by
$s_-=s_+=s$. Condition (4) of pure state is satisfied
by\cite{Serafini04}
\begin{equation}
\mathrm{det}(V)=\frac{1}{16}\label{eq02}.
\end{equation}
The resulting equation
\begin{equation}
(s^2-h_-^2)(s^2-h_+^2)=\frac{1}{16}\label{eq03}
\end{equation}
together with $V>0$ entails $s>\mathrm{max}(1/2,|h_-|,|h_+|)$.

Combining this result with the uncertainty principle for Gaussian
states,\cite{Simon94} $V+\frac{i}{2}\Omega\geq 0$, where $\Omega$
is the usual symplectic form
\begin{equation}
\Omega\equiv \left ( \begin{array}{cccc} 0 & 1 & 0 & 0 \\ -1 & 0 &
0 & 0 \\ 0 & 0 & 0 & 1 \\ 0 & 0 & -1 & 0 \end{array} \right ),
\label{eq04}
\end{equation}
gives $h_-=-h_+$, which, combined with Eq.~(\ref{eq03}) and
substituted in Eq.~(\ref{eq01}), reduces the covariance matrix to
the form
\begin{equation}
V=\half\left ( \begin{array}{cccc} \nu & 0 & \sqrt{\nu^2-1} & 0 \\
0 & \nu & 0 & -\sqrt{\nu^2-1}
\\ \sqrt{\nu^2-1} & 0 & \nu & 0 \\ 0 & -\sqrt{\nu^2-1} & 0 & \nu \end{array} \right )\label{eq05}.
\end{equation}
If one defines $\nu=\cosh(2r)$, then this Gaussian state describes
the two-mode squeezed vacuum state,\cite{Weedbrook12} also known
as an Einstein-Podolski-Rosen (EPR) state, with the squeezing
parameter $r>0$. In the number state basis the EPR state is given
by Eq.~(\ref{eq06}).

\section{QD density matrix properties for even-even measurement}
\label{app-B}
We show that the QD density matrix for an even-even measurement in
Eq.~(\ref{neweq22}) satisfies $q_{\updownarrow}\geq q_+$.
Eq.~(\ref{neweq22}) is obtained by tracing out the photons in
$|\psi_{n_1,n_2}\rangle\langle \psi_{n_1,n_2}|$, where
$|\psi_{n_1,n_2}\rangle$ is given in Eq.~(\ref{neweq21}). The
coefficient $q_\updownarrow$ is
\begin{eqnarray}
q_\updownarrow=\frac{1-\lambda^2}{2} \biggl
|P_{n_1,n_2}U_{BS}\sum\limits_{m,m'=0}^\infty
\lambda^{m+m'}\times \nonumber\\
\sin(g\sqrt{m}t)\sin(g\sqrt{m'}t)|m-1,m;m'-1,m'\rangle \biggr
|^2,\label{appeq01}
\end{eqnarray}
and $q_{+}$ is seen to have the same form except for
$|m-1,m;m'-1,m'\rangle\rightarrow |m-1,m;m',m'-1\rangle$. Up to
normalization, Eq.~(\ref{appeq01}) is the probability for the
state following $P_{n_1,n_2}$ inside the norm to be in the space
projected by $P_{n_1,n_2}$. The norm of this state and the one of
the corresponding state in $q_+$ are equal by the unitarity of
$U_{BS}$.

Using the beam splitter transformation (\ref{eq14}), we have
\begin{eqnarray}
&q_\updownarrow=\frac{1-\lambda^2}{2\sqrt{2}}\left |P_{n_1,n_2}(c_+^\dagger+d_+^\dagger)U_{BS}\sum\limits_{m,m'=0}^\infty\frac{1}{m'}\lambda^{m+m'}\times\right . \nonumber\\
&\left .
\sin(g\sqrt{m}t)\sin(g\sqrt{m'}t)|m-1,m,m'-1,m'-1\rangle\right
|^2,\nonumber\\
\label{appeq02}
\end{eqnarray}
where $q_+$ is given by the same expression with $c_+\rightarrow
c_-$ and $d_+\rightarrow d_-$. Consider the state appearing after
$(c_+^\dagger+d_+^\dagger)$ in Eq.~(\ref{appeq02}). This state
exists in both the expressions for $q_\updownarrow$ and $q_+$. Had
the operator $(c_\pm^\dagger+d_\pm^\dagger)$ not existed,
Eq.~(\ref{appeq02}) would, by virtue of the total photon number,
have given a nonzero result only for even-odd or odd-even
$(n_1,n_2)$ pairs. Since the state norm is identical in
$q_\updownarrow$ and $q_+$ both before and after the operation of
$(c_\pm^\dagger+d_\pm^\dagger)$, which adds a single photon, this
operator merely shifts probability of projection from even-odd and
odd-even pairs to even-even and odd-odd ones. Since
$q_\updownarrow$ is zero for odd-odd pairs, a given even-even pair
in it, $(n_1,n_2)$, receives all contributions from $(n_1-1,n_2)$
and $(n_1,n_2-1)$, while in $q_+$ it shares these contributions
with the odd-odd pairs $(n_1-1,n_2+1)$ and $(n_1+1,n_2-1)$. We
thus conclude that for the even-even case $q_{\updownarrow}\geq
q_+$.

\section{Time-averaged measurement probability}
\label{app-C}
We set to find the form of $C_{n_1,n_2}$, the time average of
$g_{n_1,n_2}(t)$ in Eq.~(\ref{eq26}), which is the time-dependent
part of the measurement probability. As the beam splitter
conserves photon number in each polarization, Eq.~(\ref{eq26})
reduces to
\begin{widetext}
\begin{eqnarray}
g_{n_1,n_2}(t)&=&\half\sum\limits_{k=\mathrm{max}(n_1-n_3,0)}^{\mathrm{min}(n_1,n_3)}\left [\sum\limits_{m=1}^{n_3}A_{km}\sin(g\sqrt{m}t)\sin(g\sqrt{n_3-m+1}t)\right ]^2,\label{eq27}\\
A_{km}&\equiv&\langle
k,n_1-k;n_3-k,n_3-n_1+k|U_{BS}|m-1,m;n_3-m+1,n_3-m\rangle.\label{eq28}
\end{eqnarray}
\end{widetext}
The beam splitter matrix element can be broken to a product of two
one-mode beam splitter matrix elements as
\begin{eqnarray}
&\hspace{-1.5em}\langle
m'_-,m'_+;\tilde{m}'_-,\tilde{m}'_+|U_{BS}|m_-,m_+;\tilde{m}_-,\tilde{m}_+\rangle=\nonumber\\
&\hspace{-1.5em}\langle
m'_-,\tilde{m}'_-|U_{BS,1}|m_-,\tilde{m}_-\rangle\langle
m'_+,\tilde{m}'_+|U_{BS,1}|m_+,\tilde{m}_+\rangle,\label{eq29}
\end{eqnarray}
with $U_{BS,1}$ being the one-mode beam splitter operator. By
Eq.~(\ref{eq14}) the matrix elements of this operator are
\begin{eqnarray}
&\langle k,n-k|U_{BS,1}|l,n-l\rangle=\nonumber\\
&\sqrt{\frac{k!(n-k)!}{l!(n-l)!}}2^{-n/2}\sum\limits_{q=0}^{k}{l\choose
q}{n-l\choose k-q}(-1)^{l+q}.\label{eq30}
\end{eqnarray}

The time average of a four sines product in the expansion of
Eq.~(\ref{eq27}) is nonzero only if the four frequencies of the
sines contain two identical frequency pairs. Consequently, this
time average reads
\begin{eqnarray}
&C_{n_1,n_2}=\frac{1}{8}\sum\limits_{m=1}^{n_3}(B_{mm}+B_{m,n_3-m+1})\nonumber\\
&-\frac{1}{32}B_{(n_3+1)/2,(n_3+1)/2}(1-(-1)^{n_3}),\label{eq32}
\end{eqnarray}
where
\begin{equation}
B_{mm'}\equiv
\sum\limits_{k=\mathrm{max}(n_1-n_3,0)}^{\mathrm{min}(n_1,n_3)}
A_{km}A_{km'},\label{eq33}
\end{equation}
the dependence of $A_{km}$ and $B_{mm'}$ on $(n_1,n_2)$ being
understood.

\section{Success probability for nonideal detectors}
\label{app-D}
\begin{figure}[t!]
\begin{center}
\large\flushleft{(a)}\normalsize\\
\vspace{-1em}\includegraphics[width=8.5cm]{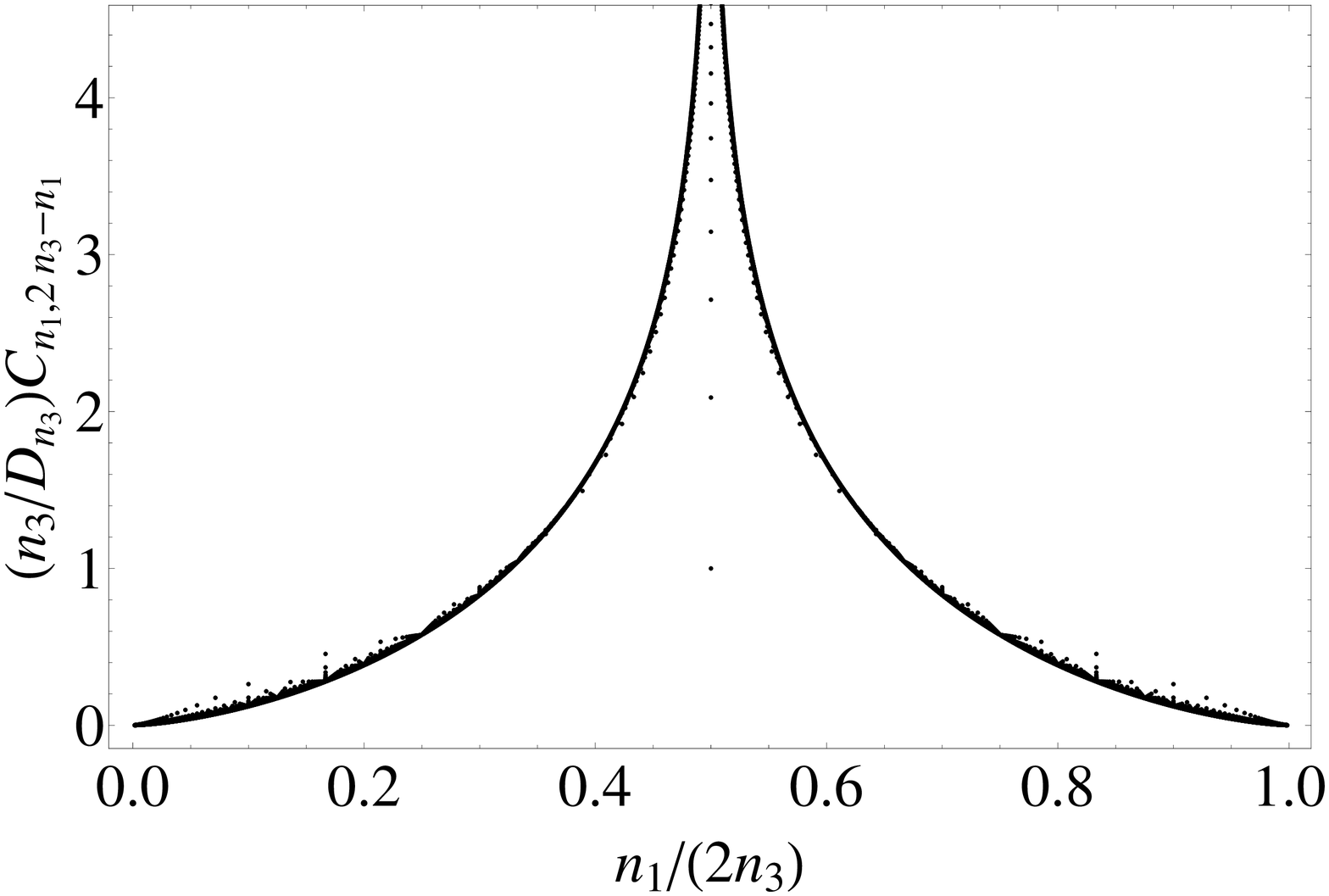}\\
\vspace{-1em}\large\flushleft{(b)}\normalsize
\end{center}
\begin{center}
\vspace{-2.1em}\includegraphics[width=8.5cm]{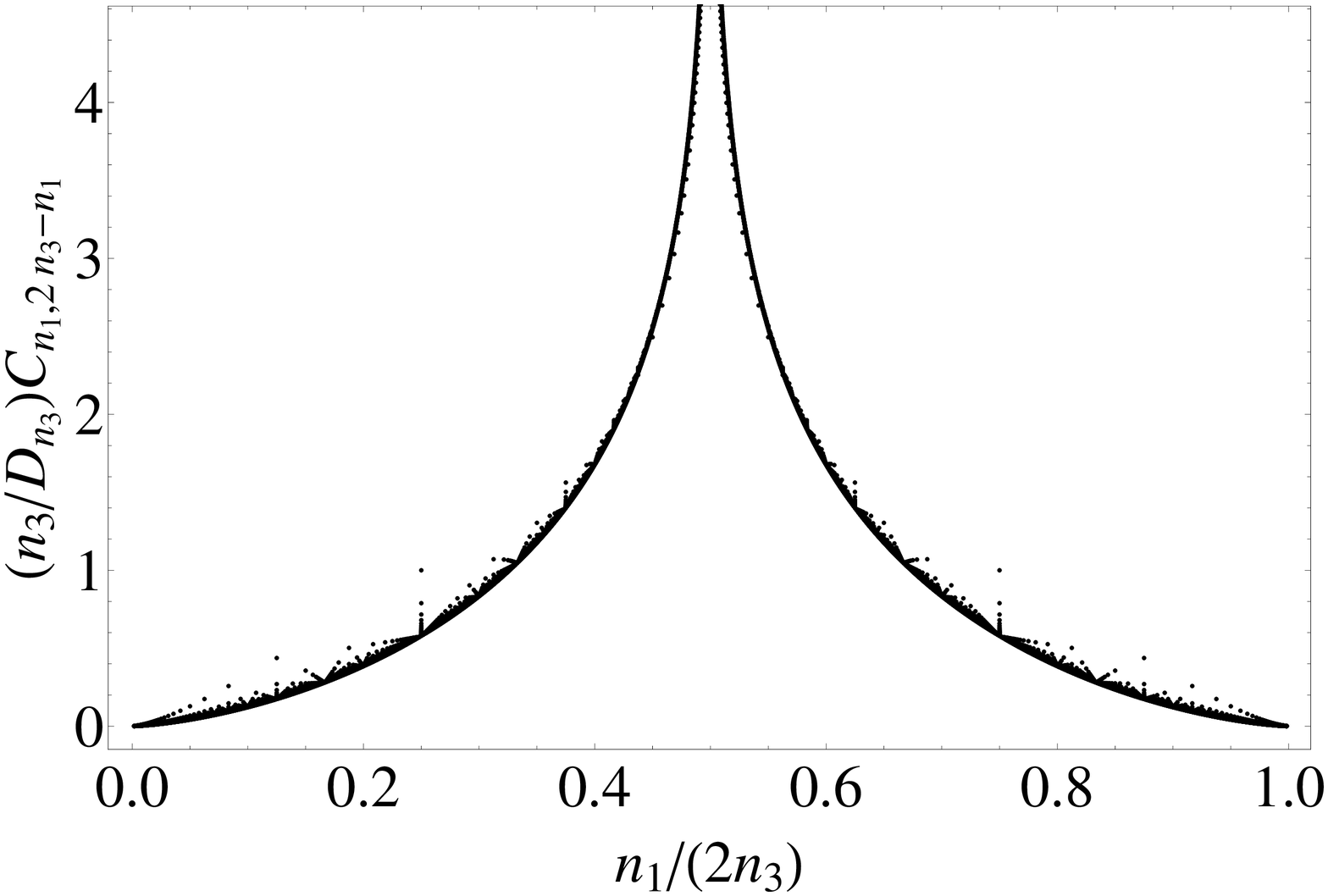}
\end{center}
\caption{\label{fig04} Normalized values of $C_{n_1,n_2}$, the
time average of the time-dependent part in the measurement
probability, for odd (a) and even (b) values of $1\leq n_3\leq
300$, where $n_3=(n_1+n_2)/2$, and $n_1$ and $n_2$ are the photon
numbers measured in the first and second detectors, respectively.
The coefficient $D_{n_3}$ is obtained by summing $C_{n_1,n_2}$
over all pairs of odd $n_1$ and $n_2$ corresponding to a given
$n_3$. The normalized values of $C_{n_1,n_2}$ exhibit the property
of \emph{data collapse} when $n_3$ is large.}
\end{figure}

We wish to find the infinite sum for the success probability with
nonideal detectors in Eq.~(\ref{eq39}). Due to the additional
$\eta$-dependent factor in Eq.~(\ref{eq39}), the sum rule of
Eq.~(\ref{eq36}) cannot be used to remove the sum over $n_1$ and
put it in closed form as in the ideal detectors case.
Approximating Eq.~(\ref{eq39}) by summing a finite number of terms
is also unsatisfactory due to the delicate convergence of this
series. Equation (\ref{eq31}) indicates that the
$\overline{m}$-dependent part of high $n_3$ summands decays to
zero very slowly relative to lower $n_3$ terms and cannot be
neglected when $\overline{m}$ is large, while $C_{n_1,n_2}$ for
these summands also does not decay with increasing $n_3$, being
always of the order of $1$ as seen from Eq.~(\ref{eq36}). In order
to find an effective approximation for the infinite sum in
Eq.~(\ref{eq39}), we first need to study the properties of
$C_{n_1,n_2}$. We use Eq.~(\ref{eq32}) to plot the normalized
$C_{n_1,n_2}$ in Fig.~\ref{fig04}, and observe that the normalized
coefficients exhibit the property of \emph{data collapse} when
$n_3$ is large. In particular, we find the standard deviation of
the $C_{n_1,n_2}/D_{n_3}$ distribution for a given $n_3$ is a
linear function of $n_3$.

With the properties of the $C_{n_1,n_2}$ coefficients in mind, we
set to find an approximation which will allow us to sum
Eq.~(\ref{eq39}). Such an approximation, inspired by
Fig.~\ref{fig04}, is the ``rooftop'' approximation, wherein
$C_{n_1,2n_3-n_1}$ for a given $n_3$ is a piecewise linear
function of the form
\begin{equation}
C_{n_1,2n_3-n_1}\approx\left \{ \begin{array}{ll}
a_{n_3}n_1+b_{n_3} \quad &\phantom{0}0\leq n_1\leq n_3\\
a_{n_3}(2n_3-n_1)+b_{n_3} \quad &n_3\leq n_1\leq 2n_3 \end{array}
\right . ,\label{eq40}
\end{equation}
where $a_{n_3}$ and $b_{n_3}$ are determined by two constraints:
Eq.~(\ref{eq36}) and the linear relation for the standard
deviation of the $C_{n_1,n_2}/D_{n_3}$ distribution. Equation
(\ref{eq40}) is substituted in Eq.~(\ref{eq39}), which can then be
exactly summed. By comparing the approximate $C_{n_1,n_2}$ with
their exact values, we find the rooftop approximation error is
lower than 2\%. We note the approximation is exact when
$\overline{m}\rightarrow \infty$, as then the dominant terms in
Eq.~(\ref{eq39}) have high $n_3$ and an essentially flat
$\eta$-dependent factor, a fact which, combined with $C_{n_1,n_2}$
still satisfying Eq.~(\ref{eq36}), leads to an exact result.

This insight leads to an improved approximation, the ``hybrid''
approximation, obtained by summing Eq.~(\ref{eq39}) up to some
high value of $n_3$, $n_{3,max}$, exactly and then writing the
rest of the sum in closed form using the rooftop approximation in
Eq.~(\ref{eq40}). With $n_{3,max}=300$, we find, by the method
noted above, that the hybrid approximation error is below 1\%. We
use this approximation to plot Fig.~\ref{fig05}.

\begin{acknowledgments}

This research was supported by U.S. Army Research Office MURI
Award No. W911NF0910406, by NSF Grant No. PHY-1104446 and by ARO
(IARPA, W911NF-08-1-0487). The authors thank D. G. Steel for
useful discussions.
\end{acknowledgments}

\section*{References}
\bibliography{GAUSSIAN}
\end{document}